\documentclass[fleqn, usenatbib]{mnras}
\usepackage{newtxtext,newtxmath}
\usepackage[T1]{fontenc}
\usepackage{ae,aecompl}
\usepackage[]{units}
\usepackage{graphicx}   
\usepackage{amsmath}    
\usepackage{amssymb}    
\usepackage{graphicx}   
\usepackage{amsmath}    
\usepackage{amssymb}    
\usepackage{layouts}
\usepackage{multicol}        
\usepackage{bm}         
\usepackage{pdflscape}  
\usepackage{verbatim}

\def\msun{M$_{\sun}$}

\title[Gaia WDs]{Gaia Reveals Evidence for Merged White Dwarfs}
\author[Kilic et al.]
{Mukremin Kilic$^{1,2}$,
N. C. Hambly$^2$,
P. Bergeron$^3$,
C. Genest-Beaulieu$^3$,
N. Rowell$^2$
\\
$^1$Department of Physics and Astronomy, University of Oklahoma, 440 W. Brooks St., Norman, OK, 73019, USA\\
$^2$Institute for Astronomy, University of Edinburgh, Royal Observatory, Blackford Hill, Edinburgh EH9 3HJ, UK\\
$^3$D\'epartement de Physique, Universit\'e de Montr\'eal, C.P. 6128, Succ. Centre-Ville, Montr\'eal, QC H3C 3J7, Canada\\
}

\date{\ \ Submitted \today \vspace{-0.5cm}}
\pubyear{2018}

\begin{document}
\label{firstpage}
\pagerange{\pageref{firstpage}--\pageref{lastpage}}
\maketitle

\begin{abstract}
We use Gaia Data Release 2 to identify 13,928 white dwarfs within 100 pc of the Sun. The exquisite astrometry
from Gaia reveals for the first time a bifurcation in the observed white dwarf sequence in both Gaia and the Sloan
Digital Sky Survey (SDSS) passbands. The latter is easily explained by a helium atmosphere white dwarf fraction
of 36\%. However, the bifurcation in the Gaia colour-magnitude diagram depends on both the atmospheric composition
and the mass distribution. We simulate theoretical colour-magnitude diagrams for single and binary white dwarfs using a population synthesis approach and demonstrate that there is a significant contribution from relatively massive white dwarfs that likely
formed through mergers. These include white dwarf remnants of main-sequence (blue stragglers) and post-main
sequence mergers. The mass distribution of the SDSS subsample, including the spectroscopically confirmed white
dwarfs, also shows this massive bump. This is the first direct detection of such a population in a volume-limited sample.
\end{abstract}


\section{Introduction}

Since the detection of the companion of Sirius in 1862, we have been able to create a complete census of
white dwarfs (WDs) only to within 13 pc of the Sun \citep{holberg08,holberg09}. This sample contains 43 stars, which corresponds to
a WD space density of $4.8 \pm 0.5 \times 10^{-3}$ pc$^{-3}$. \citet{holberg16} and \citet{limoges15} extended
the local sample to 25 and 40 pc, respectively, with the latter containing nearly 500 stars. However, the current sample of WDs
with reliable parallax measurements is limited to about 250 objects \citep{bergeron01,gianninas15,bedard17}.
The lack of distance measurements severely limits our understanding of the local WD population.

The fraction of binary WDs in the local 20 pc sample is 26\% \citep{holberg16}. \citet{toonen17} used a population
synthesis approach to explain this relatively low binary fraction compared to main-sequence stars ($\sim$50\%).
Based on the $\approx100$ stars in the 20 pc sample, they found that the low binary fraction is mostly caused by mergers
in binary systems. Their binary population synthesis models predict that $\sim56$\% of WDs come from single star evolution
in isolation, $\sim$7-23\% come from single WDs that formed as a result of mergers in binary systems, and the rest
are in (un)resolved double WD or WD + main-sequence star binaries. 
 
The European Space Agency (ESA)'s {\em Gaia} mission provides an unprecedented opportunity to
assemble the first reliable Hertzsprung-Russell diagram for nearby field WDs. The {\em Gaia}
Data Release 2 (DR2) presents $G$ passband photometry, integrated magnitudes in the blue (BP, 330-680 nm)
and red (RP, 640-1000 nm) passbands, and the five-parameter astrometric solution for more than 1.3 billion sources.
Now that {\em Gaia} Data Release 2 has increased the local WD sample size by two orders of magnitude \citep{gaia18}, we can
look for the signatures of a merger population directly in the {\em Gaia} colour-magnitude diagrams. 

Here we use a population synthesis approach to study the observed {\em Gaia} WD sequence. In \S 2 we
describe our source selection and filtering of the {\em Gaia} data, and in \S 3 we present our synthetic colour-magnitude
diagrams and discuss the He atmosphere WD fraction, and the bifurcation in the Gaia and SDSS passbands. We discuss the WD
mass distribution and the results from binary population synthesis studies in \S 4, and conclude in \S 5.
 
\section{Gaia Sample Selection}

In order to derive a clean sample within 100~pc of the Sun we follow the recommendations outlined in
\cite{lindegren18} to remove non-Gaussian outliers in colour and absolute magnitude. We made cuts on
BP, RP and parallax signal--to--noise of 10 or greater and employed the astrometric and photometric
quality cuts outlined in Appendix~C of \cite{lindegren18}.
A simple cut in $(G_{\rm BP} - G_{\rm RP}, M_{\rm G})$ space keeping only those sources fainter than
the line joining (-1,5) and (5, 25) was used to net the clearly subluminous stellar objects relative to the main sequence.
This selection is optimized for reliability rather than completeness; it keeps isolated WDs and unresolved double
WDs, but removes WD + main-sequence binaries from the sample, which make up about 20\% of the local WD sample \citep{holberg08}. The query executed on the {\em Gaia} archive is available as supplementary data; it returns 13,928 sources, which
corresponds to a space density of $3.3 \times 10^{-3}$ pc$^{-3}$. This is significantly lower than the local WD space
density of $4.5 \pm 0.4 \times 10^{-3}$ pc$^{-3}$ as measured from the Gaia 20 pc WD sample \citep{hollands18}. However, taking into
account the missing WD + main-sequence binaries in the 100 pc sample brings the space density up to within $1\sigma$ of the
estimate from the Gaia 20 pc sample. About 4.4\% of the WDs in our sample appear overluminous or undermassive
($M<0.5 M_{\odot}$, see \S 3.3); these are likely unresolved double degenerate binary systems. This fraction
is comparable to the 6\% double degenerate fraction found in the 20 pc sample \citep{holberg08}.

\section{Analysis}

\subsection{Single Star Population Synthesis}

\begin{figure}
\vspace{-0.5in}
\includegraphics[width=\columnwidth]{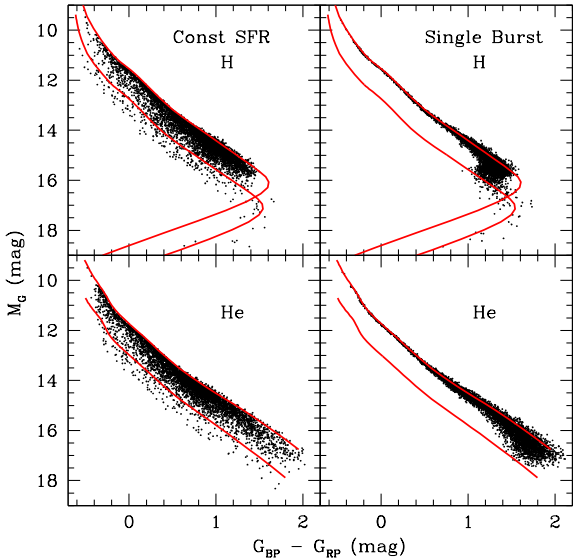}
\vspace{-0.9in}
\caption{Synthetic colour-magnitude diagrams for H (top panels) and He (bottom panels) atmosphere WDs
for a 10 Gyr old disc population and assuming a constant star formation rate (left panels) or an initial single burst of star
formation (right panels). Each panel displays 5,000 WDs that evolved in isolation. The solid lines show
the appropriate model sequences for 0.5 and $1.0 M_{\odot}$ WDs for each composition.}
\label{fig:single}
\end{figure}

The canonical age estimate for the thin and thick discs are about 8-10 Gyr \citep{leggett98,kilic17}. To model the disc
populations, we assume a constant star formation rate and generate 100 main-sequence stars at every 1 Myr time step with masses randomly drawn between 0.8 and 8\msun\
from a Salpeter mass function \citep{salpeter55}. We assume solar metallicity and use the main-sequence + giant-branch lifetimes from \citet{hurley00} to decide if a star evolves into a WD within the
model age. We calculate the final WD masses based on the initial-final mass relation (IFMR) from \citet{kalirai08}. Given the total age
and the main-sequence lifetime of each star, we estimate its WD cooling age. Combined with the final WD
mass, the cooling age enables us to calculate the expected colours and magnitudes \citep{bergeron95,holberg06} of a given star in the
{\em Gaia} passbands. Here we use the Gaia DR2 passbands that were used to calculate the published DR2 values \citep{evans18}. 

To explore the effects of the star formation history of the disc on our simulations, we calculate another set of models where
we assume a single burst of star formation that lasts 1 Gyr. We also assume a metallicity of [Fe/H] = $-0.7$ \citep{ivezic08},
which is appropriate for an old population like the thick disc. We use the evolutionary lifetimes from \citet{hurley00} for that metallicity, and follow the same procedure as above to estimate the final WD mass,
cooling age,  {\em Gaia} magnitudes and colours. 

Figure \ref{fig:single} shows the synthetic colour-magnitude diagrams for 10 Gyr old disc WD
populations assuming either a constant star formation rate or a single burst. 
A constant star formation rate leads to the formation of WDs with a variety of
masses (and therefore absolute magnitudes) at any given age. On the other hand, all of the intermediate-mass stars in
the 10 Gyr old single burst disc population have already evolved into cool WDs and this population is currently
forming only $M\sim0.5 M_{\odot}$ WDs, creating a narrow band of stars at the bright end
of the WD sequence.

\subsection{The He atmosphere WD Fraction}

\begin{figure}
\vspace{-0.45in}
\includegraphics[width=3in]{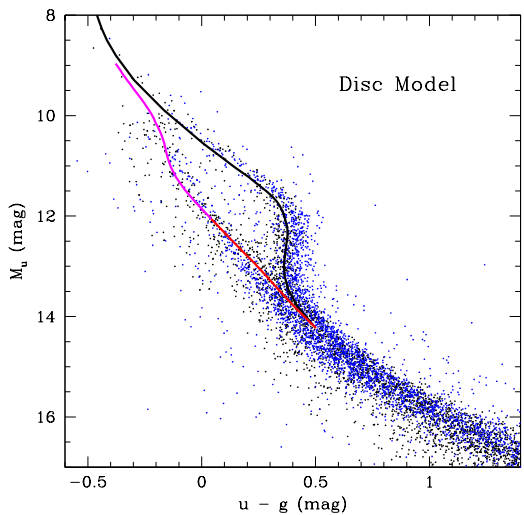}
\vspace{-0.8in}
\caption{SDSS WDs within 100 pc of the Sun (blue points) and our synthetic model for a 10 Gyr
old disc WD population (black points). Model sequences for pure H (black line, $T_{\rm eff}\geq7000$ K) and
pure He atmosphere (magenta for $T_{\rm eff}>11,000$ K and red between 7000-11,000 K) are also shown. He lines
disappear below about 11,000 K. Hence, pure He atmosphere WDs below that temperature would be classified as
DC spectral type.}
\label{fig:he}
\end{figure}

Figure \ref{fig:he} shows a colour-magnitude diagram for the SDSS WDs in the 100 pc sample. 
There is a clear split of the WD sequence into H and He atmosphere WDs in the narrower
SDSS  passbands \citep{gaia18}, which are more sensitive to the atmospheric composition. H-rich WDs
suffer from the Balmer jump. Their ultraviolet fluxes are suppressed  compared to their He-rich counterparts,
and they have redder $u-g$ colours. The bifurcation due to atmospheric composition is visible for $M_u<14$ mag,
which corresponds to $T_{\rm eff}\approx 7000$ K.

To estimate the He atmosphere WD fraction, we create synthetic colour magnitude
diagrams for a variety of ages (6-10 Gyr) and He atmosphere WD fractions (10-50\%). We then divide the observed
colour-magnitude diagram into 80 boxes within the range $M_u=10$-14 and $u-g=-0.4$ to +0.6 mag, and compare the
relative number densities of each model with the data to find the best model that reproduces the observed number densities.
This is similar to the analysis done on the globular cluster WD sequences of M4, NGC 6397, and 47 Tuc
\citep{hansen07,hansen13}.  We then perform a Monte Carlo analysis where we replace the observed magnitudes
with $G + g \sigma_G$ where $g$ is a Gaussian deviate with zero mean and unit variance. For each set of the modified
data, we find the best-fitting model, and take the range in parameters that encompasses 68\% of the probability distribution
function as the $1\sigma$ uncertainties. We constrain the fraction of He atmosphere WDs to $36 \pm 2$\% in this colour and
magnitude range.

The ratio of He- to H-atmosphere WDs is strongly temperature dependent because of convective mixing
and dilution. The evolution of the atmospheric composition due to these effects is not well understood \citep{bergeron01,chen12}.
Hence, our results are only valid in the $M_u=10$-14 mag range. Further insight into the He atmosphere WD fraction of the 100 pc 
population will have to await a detailed model atmosphere analysis that relies on spectroscopy and near-infrared photometry to
distinguish between H and He atmosphere WDs.

\subsection{The Bifurcation in the Gaia Passbands}

\begin{figure}
\includegraphics[width=\columnwidth]{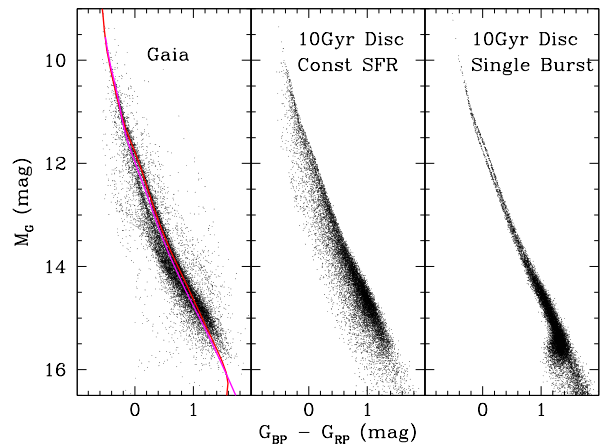}
\vspace{-0.2in}
\caption{Observed (left panel) and synthetic colour-magnitude diagrams for the 100 pc WD sample in the Gaia passbands.
Model sequences for $0.6 M_{\odot}$ H (red) and He (magenta) atmosphere WDs are shown in the left panel. 
The middle and right panels show the predicted sequences for a 10 Gyr old disc population assuming either a constant star
formation rate or a single burst of star formation. The bifurcation in the Gaia data is not matched by our simulations based
on single star population synthesis and pure H and pure He WD model atmospheres.}
\label{fig:gaia}
\end{figure}

Figure \ref{fig:gaia} shows the Gaia 100 pc WD sequence (left panel) and our synthetic colour-magnitude diagrams
based on single star population synthesis and a He atmosphere WD fraction of 36\%. As noted in \citet{gaia18}, there
is a clear bifurcation of the observed
sequence below $G=12$ mag, but this is not reproduced by our simulations including a large fraction of pure He atmosphere WDs. 
To explore the source of this bifurcation, we cross-matched the SDSS subsample with the Montreal White Dwarf Database \citep{dufour17}
and found 800 spectroscopically confirmed WDs with $T_{\rm eff} \geq 6000$ K. Since H$\alpha$ is still visible at these temperatures
for H atmosphere WDs, this temperature selection enables us to define relatively clean samples of DA and non-DA white dwarfs.

\begin{figure}
\vspace{-0.35in}
\includegraphics[width=\columnwidth]{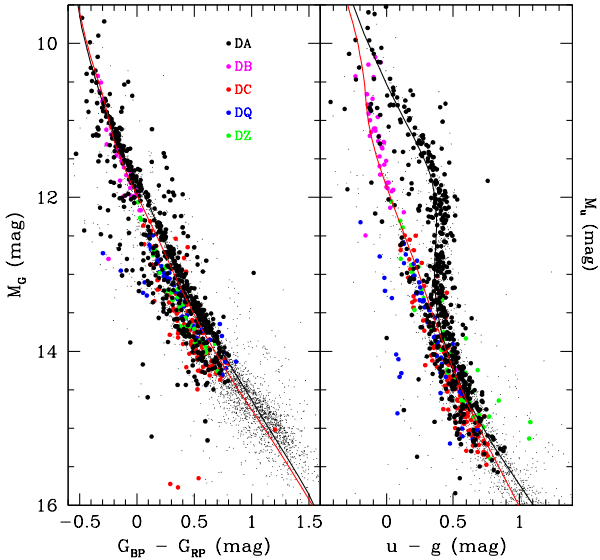}
\vspace{-0.9in}
\caption{Gaia (left panel) and SDSS (right panel) colour-magnitude diagrams for the spectroscopically confirmed white
dwarfs with $T_{\rm eff}\geq 6000$ K in the SDSS subsample. Model sequences for H (black line) and He (red line) atmosphere
WDs are also shown.}
\label{fig:spec}
\end{figure}

Figure \ref{fig:spec} shows Gaia and SDSS colour magnitude diagrams for the spectroscopically confirmed WDs in the SDSS subsample.
DA and DB WDs follow the pure H or pure He atmosphere model predictions relatively well in both the SDSS and Gaia passbands.
However, most of the DC, DQ, and DZ WDs fall to the left of the $0.6 M_{\odot}$ pure He atmosphere model predictions in the Gaia colour-magnitude 
diagram (left panel). Since DQ and DZ WDs show absorption features that are not included in the pure He atmosphere models,
their bluer colours are not surprising, and therefore atmospheric composition can partly explain the bifurcation seen in the Gaia bands.
However, DC WDs should follow the pure He atmosphere models, unless they are, on average, more massive than expected. Furthermore, there
is a large number of spectroscopically confirmed DA WDs that also fall on the lower sequence observed in the Gaia passbands.
Hence, the bifurcation in the Gaia passbands is not due to atmospheric composition alone; the mass distribution also plays a role.

\begin{figure}
\vspace{-0.35in}
\includegraphics[width=3in]{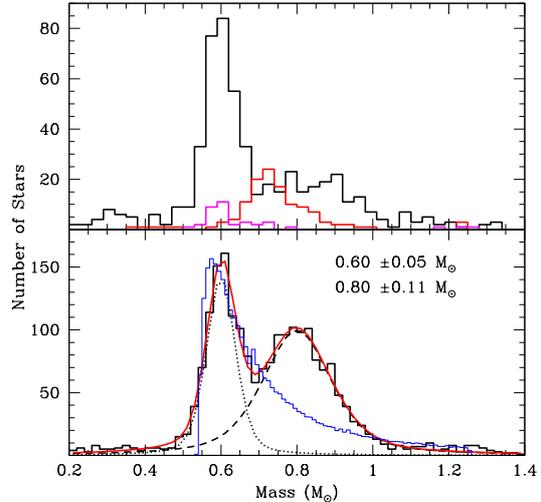}
\vspace{-0.9in}
\caption{{\it Top panel:} The mass distributions for the spectroscopically confirmed DA (black), DB (magenta), and DC (red)
WDs shown in Figure \ref{fig:spec}. {\it Bottom panel:} The mass distribution for the 100 pc WDs
with $T_{\rm eff}\geq 6000$ K based on pure H atmosphere model fits to their $ugriz$ photometry and Gaia parallaxes.
The dotted and dashed lines show the best-fit pseudo-Voigt profiles to the two populations with mean masses of
0.60 and  $0.80 M_{\odot}$,  respectively, and the red line shows the combined fit.
The blue line shows the predicted mass distribution for a disc population with a constant star formation rate.}
\label{fig:ugriz}
\end{figure}

The top panel in Figure \ref{fig:ugriz} shows the mass distributions of the DA, DB, and DC WDs in the SDSS subsample using
pure H (for DA) and pure He (for DB and DC spectral types) atmosphere models. The DA mass distribution
peaks at $0.6 M_{\odot}$ but displays a significant contribution from massive WDs.  The DC WDs are also relatively massive.
 \citet{liebert05} discuss the volume effect for detecting different mass WDs in magnitude-limited surveys. Massive WDs are intrinsically
 fainter, hence the effective survey volume for such WDs is significantly smaller. Hence, the total number of massive WDs are
 significantly underestimated in magnitude-limited spectroscopic surveys like the SDSS.

The bottom panel in Figure \ref{fig:ugriz} shows the mass distribution of all 100 pc SDSS WDs ($\approx$80\%
lack spectroscopy) with $T_{\rm eff}\geq 6000$ K
based on our pure H atmosphere model fits to the $ugriz$ photometry and Gaia parallaxes. Clearly the mass distribution of
the WDs in the solar neighborhood is bimodal with two significant peaks at 0.6 and $0.8 M_{\odot}$, respectively. 
Using pure He atmosphere models gives similar results: if we take the high mass objects with $M=0.7-1 M_{\odot}$ 
and fit them with He atmosphere models, 65\% of the stars still remain in the massive WD sample. 

The blue line in the bottom panel shows the predicted mass distribution for a disc population of single WDs. Normalizing
the disc model to match the observed mass distribution in the 0.5-$0.7 M_{\odot}$ range, we find that the secondary peak
at $0.8 M_{\odot}$ contributes 25\% of the stars in the  0.7-$1 M_{\odot}$ range. We use the spectroscopically confirmed
WDs to estimate the contamination rate of this massive bump from non-DA WDs. There are
163 DA, 32 DB, 109 DC, 46 DQ, and 31 DZ WDs for which our H atmosphere model fits place them in the massive bump.
\citet{kleinman13} present spectroscopic fits to 99 of these DAs, which have a median $\log{g}=8.43$. Hence, $\geq43$\% of the
spectroscopically confirmed WDs in the secondary peak are bonafide massive DA WDs, implying that $\sim$11\% of the
WDs in the 100 pc sample are massive.
                                  
In order to verify that this bimodal mass distribution is consistent with the SDSS colour-magnitude diagrams, we fit the Gaia $G$
and $G_{\rm BP} - G_{\rm RP}$  photometry and parallaxes of our 100 pc WD sample with pure H and pure He
atmosphere models. Given the estimated masses and cooling ages for each WD, we then calculate their predicted colours
and magnitudes in the SDSS passbands. We add random noise to their predicted magnitudes based on the SDSS
observations. We call these simulations ``the Gaia model''.

\begin{figure}
\includegraphics[width=\columnwidth]{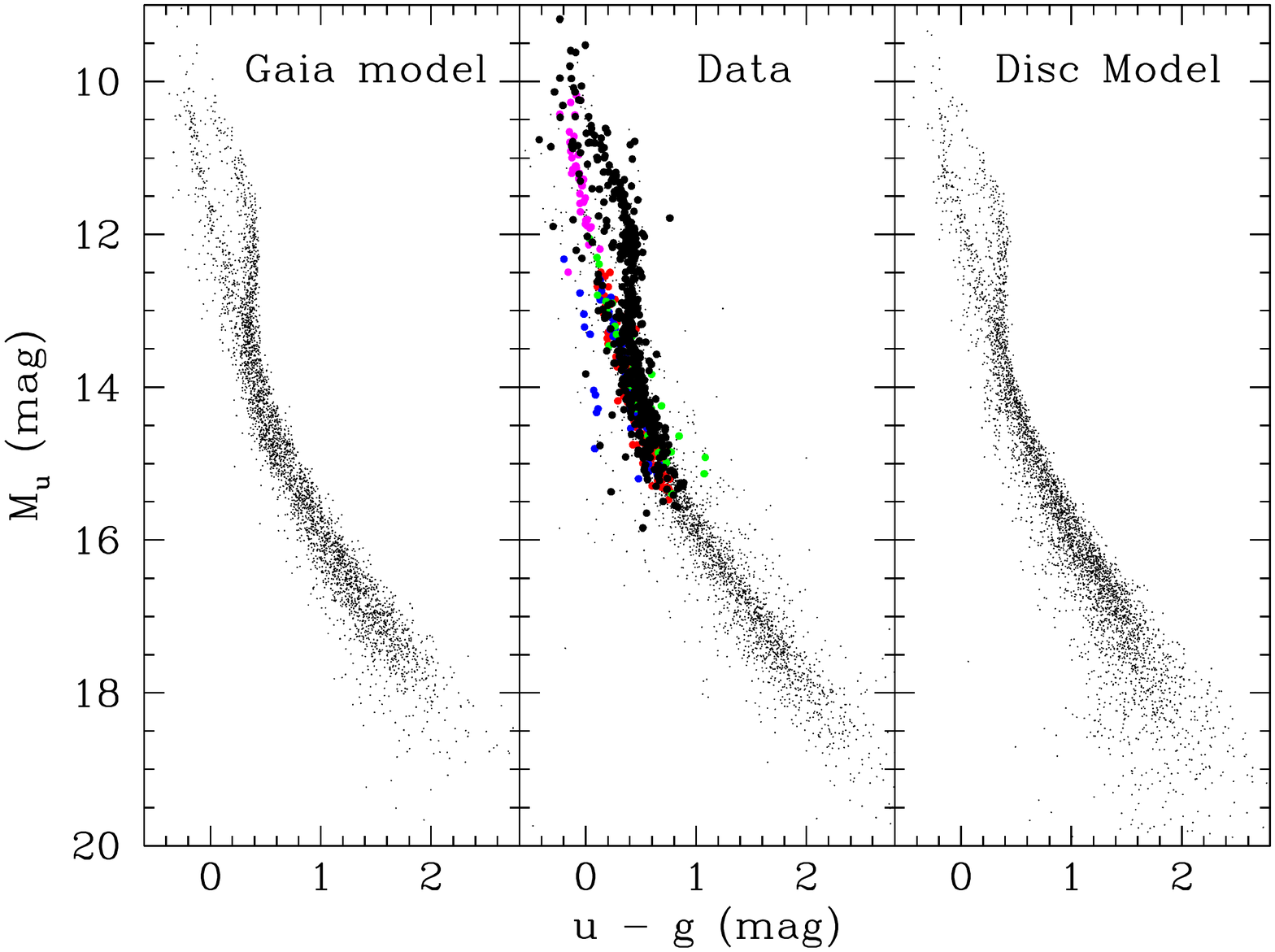}
\includegraphics[width=\columnwidth]{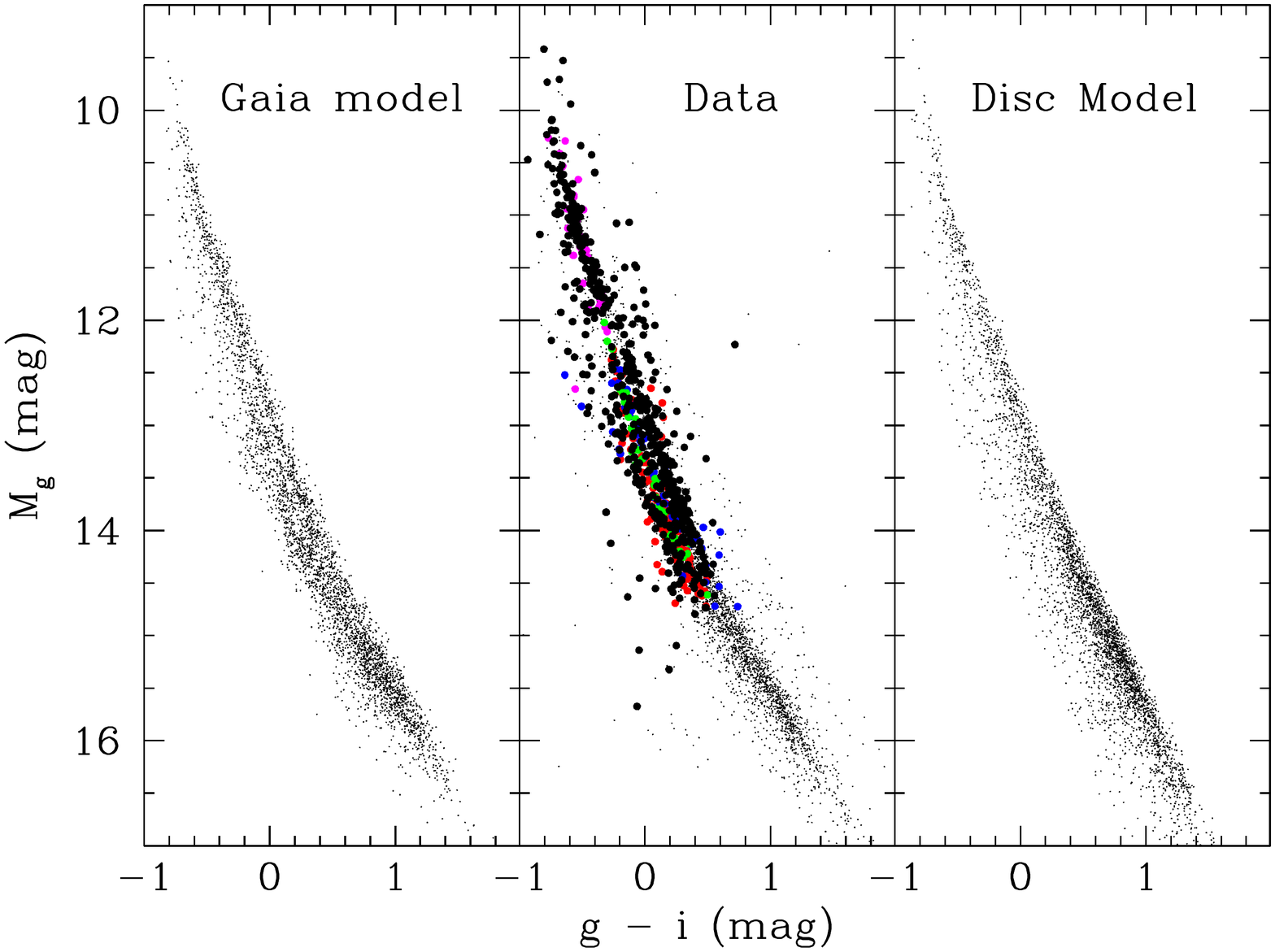}
\vspace{-0.25in}
\caption{Color-magnitude diagrams in the SDSS passbands. The left panels show the predicted colours of our 100 pc sample
of WDs based on the masses and cooling ages derived from Gaia photometry and parallaxes. The middle panels
show the observed WD sequence and the spectroscopically confirmed WDs from Figure \ref{fig:spec}, and the right panels show the synthetic sequences for a 10 Gyr old disc population of single stars with a smooth mass distribution.}
\label{fig:colour}
\vspace{-0.1in}
\end{figure}

Figure \ref{fig:colour} shows two sets of SDSS color-magnitude diagrams. The left panels show the predicted SDSS
photometry of our Gaia sample, based on the masses and cooling ages derived from Gaia photometry and
parallaxes (the Gaia model). The middle panels show the observed sample of 4016 WDs with SDSS photometry,
and the right panels show our synthetic colour-magnitude diagrams for single WDs in a 10 Gyr old disc. For a fair
comparison, we randomly select 4016 stars from the Gaia and the disc models.
This figure demonstrates that both models agree fairly well with the observed WD
sequence in the $M_u$ versus $u-g$ diagram. They both match the bifurcation at the bright end due to the atmospheric
composition, but any potential bifurcation at fainter magnitudes is lost in the noise. Hence, given the relatively
large photometric errors in the $u-$band, it is easy to hide a bifurcated mass distribution (as in our Gaia model predictions)
in this colour-magnitude diagram.

The observed WD sequence in the $M_g$ versus $g-i$ colour-magnitude diagram shows a slight
bifurcation between $M_g = 12$-14 mag. This is similar to the bifurcation seen in the Gaia passbands, $G$ versus $BP-RP$.
The disc model, which includes a 36\% He atmosphere WD fraction and a smooth mass distribution, does not match this bifurcation.
On the other hand, the Gaia model shows a similar bifurcation in the same magnitude range as the observations, though the
bifurcation is somewhat lost in the noise. Hence, there is evidence for a mass bifurcation even in the SDSS $M_g$ versus
$g-i$ colour-magnitude diagram, but the evidence is not as significant due to the large errors in the SDSS observations.

\section{Discussion}

There is considerable interest in understanding the field WD mass distribution. \citet{tremblay16} used 97 stars
from the local 20 pc sample and 715 stars from the Sloan Digital Sky Survey spectroscopy sample to study the field
WD mass distribution and concluded that there is no evidence of a population of double WD mergers in the observed
mass distribution. \citet{hollands18} revisited the 20 pc sample using Gaia parallaxes, and found a mass distribution
in agreement with the \citet{tremblay16} results. However, given the small number statistics, their mass distribution can
still be fit with a secondary peak at $0.8 M_{\odot}$ as long as the fraction of massive WDs in their sample is $<30$\%
(M. Hollands 2018, private communication). This is consistent with \citet{liebert05}, who
attribute a high mass bump seen in the Palomar-Green survey WD sample to a 15\% contribution from merged
WDs. Similarly, \citet{limoges15} find an excess of massive WDs at low temperatures in the 40 pc local
WD sample. \citet{maoz18} discuss potential reasons why massive WDs might have been excluded from the
20 pc sample and the \citet{tremblay16} study, and uses two different radial velocity surveys to argue that 8.5-11 \% of all
WDs formed through mergers.

We use a binary population synthesis approach to study the frequency of mergers. We create 100,000 main-sequence
binaries with primary masses randomly drawn between 0.8 and 8\msun\ from a Salpeter mass function and the secondary
masses drawn from a uniform mass ratio distribution between 0 and 1 \citep{duchene13}. As in \citet{toonen17}, we
assume a Lognormal period distribution with $\log{P}=5.03 \pm 2.28$ d and up to $\log{P}=10$ d. We assume
a uniform eccentricity distribution between 0 and 1, except for the closest binaries with $P<12$ days for which we assume $e=0$
\citep{raghavan10}.  We follow the evolution of these systems over 10 Gyr using the binary-star evolution (BSE) algorithm of 
\citet{hurley02}. By 10 Gyr, about 15\% of the 100,000 binaries merge either on the main-sequence or post-main sequence and
evolve into single WDs, and another 19\% form double WDs.  \citet{toonen17} demonstrated that the
numbers of merged and double WDs are sensitive to the input assumptions and that
the number of merged systems heavily depends on the initial period distribution, affecting their space density by
a factor of 2.
 
For comparison, a single burst 10 Gyr old disc population converts 77\% of its main-sequence stars into WDs. Hence,
for a main-sequence binary fraction of 50\%, a sample of 100 initial systems would produce 38.5 single disc WDs,
7.5 single WDs from mergers, and 9.5 double WDs; the fraction of single WDs that form through
mergers in binary systems is $\approx$14\%, with a factor of 2 uncertainty. In addition, the merger products are predicted to have
$M= 0.74 \pm 0.19 M_{\odot}$. Hence, a combination of single WDs that evolve in isolation in a 10 Gyr old disc and single 
WDs that form through mergers in binary systems can naturally explain both the number and mass distribution of the 100 pc
Gaia WD sample. 

\citet{elbadry18} present an alternative explanation for the large numbers of massive WDs by fine-tuning the IFMR.
Their piece-wise fit has 8 free parameters, and it requires an IFMR that is significantly flatter in the initial mass range
3.5-5.5 $M_{\odot}$ compared to the IFMR derived from open cluster WDs \citep{kalirai08,williams09}. Even though,
a flattened IFMR is plausible, it cannot explain the large difference between the binary fractions of main-sequence
stars (50\%) and their descendant WDs \citep[26\%,][]{holberg16}.

\section{Conclusions}

We use the local 100 pc sample of WDs to demonstrate that a significant fraction of the single WDs in
the solar neighborhood are massive. Thanks to the exquisite astrometry from Gaia, this population reveals
itself through a bifurcation in mass (and absolute magnitude) in the Gaia colour-magnitude
diagrams, which are also affected by the atmospheric composition. Our single and binary population synthesis
calculations show that the overall properties of the massive WDs are consistent with formation through mergers
in binary sytems.

\section*{Acknowledgements}

This work is supported in part by the NASA grant NNX14AF65G, the NSERC Canada, and by the Fund
FRQ-NT (Qu\'ebec). Kilic thanks the University of Edinburgh's Institute for Astronomy Wide Field Astronomy
Unit staff for their hospitality during his sabbatical visit.
This work presents results from the European Space Agency (ESA) space mission Gaia.
Gaia data are being processed by the Gaia Data Processing and Analysis Consortium (DPAC).
Funding for the DPAC is provided by national institutions, in particular the institutions
participating in the Gaia MultiLateral Agreement (MLA).
The Gaia mission website is https://www.cosmos.esa.int/gaia.
The Gaia archive website is https://archives.esac.esa.int/gaia.

\bibliographystyle{mnras}
\bibliography{master}

\bsp
\label{lastpage}

\end{document}